\documentstyle{article}
\begin{document}

\begin{center}
\begin{Large}
\begin{bf}

A search for the associations of distant \\

radio-bright quasars with Abell clusters:\\

an effect of gravitational amplification bias ?\\
\end{bf}
\end{Large}

\begin{Large}

\bigskip
\bigskip
\bigskip
\bigskip
\bigskip
\bigskip

 Xiang-Ping Wu$^{1,2}$ and Jinlin Han$^{1}$\\

\end{Large}

\begin{large}
\bigskip
\bigskip

{\noindent}$^{1}$Beijing Astronomical Observatory,\\
 Chinese Academy of Sciences,\\
Beijing 100080, People's Republic of China\\

\bigskip
\bigskip

$^{2}$DAEC, Observatoire de Paris-Meudon,\\
92195 Meudon Principal Cedex,  France\\

\end{large}
\end{center}

\bigskip
\bigskip
\bigskip
\bigskip
\bigskip
\bigskip
\bigskip
\bigskip

\begin{large}
\begin{center}
Accepted for Publication in:\\

\bigskip

{\bf Monthly Notices\\
of the Royal Astronomical Society}

\bigskip
\bigskip
\bigskip
\bigskip
\bigskip
\bigskip
\bigskip
\end{center}
\end{large}

{\noindent}------------------------------------------------------------------\\
e-mail: wxp\%melamb@mesiob.obspm.fr ~~or~~wxp@bao01.bao.ac.cn

\newpage
\begin{center}
\begin{Large}
\begin{bf}
A search for the associations of distant \\
radio-bright quasars with Abell clusters: \\
an effect of gravitational amplification bias ?\\
\end{bf}
\end{Large}

\bigskip
\bigskip

\begin{Large}

Xiang-Ping Wu$^{1,2}$  and  Jinlin Han$^{1}$ \\

\bigskip

($^{1}$Beijing Astronomical Observatory, Beijing 100080 - China)\\
($^{2}$DAEC, Observatoire de Paris-Meudon, Meudon 92195 - France)\\

\bigskip
\bigskip
\bigskip
\bigskip

{\underline{\bf ABSTRACT}}\\
\end{Large}
\end{center}

The theory predicts that the effect of gravitational lensing by the matter
associated with clusters of galaxies can magnify background sources,
leading to  an enhancement of source number density around foreground
clusters of galaxies.  We conduct a search for the associations  of distant
radio-bright quasars with Abell clusters using the 1-Jy and 2-Jy all-sky
catalogs. Statistics turns to be  very poor on the basis of the 1-Jy sample,
which shows no correlations between the distant radio quasars
with the foreground Abell clusters above $1\sigma$ level.
However, an apparent association ($>1\sigma$) of the 2-Jy radio sources
with the foreground Abell clusters has been detected on scale of
$\sim20$ arcminutes. We point out that this enhancement is unlikely to be
produced by the statistical gravitational lensing hypothesis utilizing the
matter associated with a population of isolated clusters unless
(1)their velocity dispersion is a few times larger than the presently
adopted value and/or (2)the intrinsic counts of the radio-bright sources
with flux have a steeper slope than the presently observed ones.
This indicates that (1)the observed associations, if real,
can be the integrated result of all the matter along the line of sight
to the distant quasars, namely, the weak lensing effect of clusters of
galaxies that trace large-scale structures of the universe,
and/or (2)the number counts of the radio-bright  sources have been
seriously contaminated by lensing.\\

\bigskip
\bigskip
\bigskip
\bigskip
{\noindent}-----------------------------------------------------------------------\\
{\noindent}{\bf Key Words:} ~~gravitational lensing -- quasars: general
                            -- radio sources: general -- galaxies: clustering\\

\newpage
\begin{Large}
\begin{center}
{\bf 1. Introduction}\\
\end{center}
\end{Large}
\bigskip

One of the important consequences of statistical gravitational lensing,
as first realized by  Gott \& Gunn (1974) before the discovery of
the first lensed quasar pair,   is that high-redshift
quasars may associate with low-redshift galaxies because flux-limited
samples can pick up the intrinsically fainter quasars magnified by the
effect of
gravitational lensing of foreground galaxies, i.e., amplification
bias.  Since the early statistical report of the observational evidence on the
presence of associations of the optically selected quasars with galaxies
(Tyson, 1986; Webster et al., 1988), the subsequent work has  been devoted
to the confirmations of such associations (see Wu (1994) for a summary)
including in different wavelengths.
Among these, Fugmann (1988; 1990) found an association of relatively
bright galaxies with distant radio quasars, and Bartelmann \&
Schneider (1993c) claimed a correlation between quasars and IRAS galaxies.
No matter whether these associations relate to foreground galaxies alone or
to even large-scale structures (Bartelmann \& Schneider, 1993a,b,c),
the effect of gravitational lensing has been found to be the natural
explanations so far.\\

Recently, Wu \& Hammer (1994) have theoretically investigated if
 distant objects can be associated with foreground galaxy clusters
through the mechanism of gravitational amplification bias. They
have concluded that it is possible nowadays to observe these associations
similar to the quasar-galaxy ones. Indeed,   several observations have
indicated that foreground clusters of galaxies can act as strong or week
lenses for distant objects, such as  galaxies and quasars. Giant luminous
arcs provide a strong evidence on the very efficient lenses of clusters
of galaxies  for background galaxies
(Lynds \& Petrosian, 1986; Soucail et al., 1987; Wu \& Hammer,
1993), while two quasars have been found to be due to the week
gravitational lensing by the matter associated with clusters of galaxies
(Stickel \& K\"uhr, 1992; Bonnet et al., 1993).
In fact, a few years ago some of the high-redshift
galaxies in the 3CR were already shown to be magnified by
foreground clusters of galaxies (Hammer \& Le F\`evre, 1990, references
therein).  During the preparation of this paper we learnt that a statistical
significant overdensity of high redshift quasars in the directions of
foreground galaxy clusters has been also detected (Rodrigues-Williams \&
Hogan 1993). \\

On the other hands, in order to reach a statistically significant enhancement
by gravitational lensing, source counts should exhibit a very steep
slope with flux.  Radio-bright sources then provide the ideal sample to
such a purpose.  The least-squares fit to a power-law using the data of the
$S_{2.7{\rm GHz}}\geq0.1$ Jy sample, the $S_{2.7{\rm GHz}}\geq1.5$ Jy sample
and the $S_{2.7{\rm GHz}}\geq2$ Jy sample (Dunlop \& Peacock, 1990) gives
the surface number density of radio sources $N(>S)\sim S^{-1.62}$.
The enhancement factor induced by amplification bias, independent of
the specific lensing models, can be evaluated through
$q=[N(>S/A)/N(>S)](1/A)$ (Narayan, 1989),
where $A$ is the magnification. Fig.1 shows the variation of $q$ with
$A$ for radio-bright sources of $S_{2.7{\rm GHz}}\geq1$ Jy.
The full variations of $q$ with $S$ and $A$ can be found in Wu \& Hammer
(1994). It turns out that a large enhancement of radio-bright sources
around the foreground lensing objects
can be expected at moderate magnification.\\

The purpose of this paper is to search for the possible associations
of foreground clusters of galaxies with distant radio-bright  quasars.
A galaxy cluster alone can act as a strong lens, as was shown above.
Moreover, if clusters of galaxies trace the large-scale structure
of the universe, distant quasars may be further affected by the weak
lensing effect of large-scale inhomogeneities (Bartelmann \& Schneider,
1993a,b,c), enhancing the chance of observing the associations.
The samples of distant radio-bright quasars are extracted from
the flux-limited radio source surveys of K\"uhr et al. (1981) and
of Wall \& Peacock (1985), and Abell clusters of galaxies are chosen for
the lensing objects. As a comparison,
we have noticed that Rodrigues-Williams \& Hogan (1993)
used the Large Bright Quasar Survey as the source sample and the
small Zwicky clusters as the lensing objects. \\

\begin{Large}
\begin{center}
{\bf 2. Sample selections}\\
\end{center}
\end{Large}
\bigskip

The all-sky catalog of 4073 Abell rich clusters of galaxies (Abell,
Corwin \& Olowin, 1989) constitutes the well-defined sample  at relatively
lower redshifts, nearly completed to $z<0.2$. With the available redshift
measurements for 121 clusters (Struble \& Rood, 1991),
the most have $z<0.2$ and the maximum reaches  $0.4$.
Therefore, Abell clusters can be regarded as foreground lensing objects
for the well separated distant sources.\\

We firstly choose the radio-bright sources from the K\"uhr et al. (1981) 1-Jy
catalog selected at 5 GHz, covering the whole sky of $|b^{ll}|>10^o$.
We have added the newly measured redshifts from the recent observations
(Stickel, k\"uhr \& Fried, 1989, 1993;  Stickel \& k\"uhr, 1993a,b,c;
Hewitt \& Burbidge, 1993). The update catalog contains 518 sources and
$90\%$ of them have spectroscopically determined redshifts.  To ensure
that the sources are distant enough to act as the lensed targets and are
not cluster numbers that associate physically with the Abell clusters
(Robertson \& Roach, 1990; Brown \& Burns, 1991; Unewisse \& Hunstead, 1991),
we  restrict the redshifts of all the radio sources to be larger than
0.5. This reduces the number of radio sources to 224, in which
we have excluded the quasars without measured redshifts and the
galaxies due to their scarcity.\\

The second sample of radio-bright quasars is extracted
from  the 2-Jy all-sky survey of Wall \& Peacock (1985) at 2.7 GHz.
This sample covers the same area of the sky as the 1-Jy one and
comprises 233 sources. With our redshift limit of $z>0.5$, the
subsequent sample contains 103 sources, among which 62 quasars have
spectroscopically measured redshifts while the estimated redshifts
have been given for the rest using the optical magnitude -- redshift
relation.   We have included the estimated redshift quasars in order
to reduce the statistical fluctuations arising from the small number of
sources. \\

\begin{Large}
\begin{center}
{\bf 3. Statistical results}\\
\end{center}
\end{Large}
\bigskip

We search for the number excess of Abell clusters in the vicinities
of background quasars. The search range is a circle of radius of
$\theta$ centered on each of the quasars of the subsamples.  We
count the clusters of galaxies that locate within the search radius,
$N(<\theta$), and compute the
surface number density of Abell clusters over the search area using
$s(<\theta)=N(<\theta)/(n\pi\theta^2)$, $n$ denoting the total
number of quasars in the subsamples.
The enhancement factor is finally $q(\theta)=s(<\theta)/s_o$, where
$s_o$ is the mean surface number density of Abell clusters.
Note that the center  positions of Abell clusters have an
uncertainty of $\sim 1$ arcminute and it then turns
out that a search radius less than 1 arcminute is meaningless.\\

Table 1 lists one of the results, the detected Abell clusters within a
radius of  $\theta\leq0.3^o$ around quasars from the 224 1-Jy sample.
The same  method is used over different search ranges in both the 1-Jy
and 2-Jy samples. The number of detected Abell clusters has been
given in Table 2, and  the variations of enhancement factors normalized
at $\theta=3^o$ with the search radii are shown in Fig.2 and
Fig.3 for the 1-Jy and 2-Jy samples, respectively.
The error bar is the standard deviation in each measurement.   \\


\begin{Large}
\begin{center}
{\bf 4. Discussion}\\
\end{center}
\end{Large}

Unfortunately, significant number excess of Abell clusters in the directions
of background radio-bright quasars have not been detected  in both
the 1-Jy and the 2-Jy samples, in comparison with the recent
claim (Rodrigues-Williams \& Hogan, 1993) that there exist strong
associations between the foreground galaxy clusters with the background
bright quasars. This might be due partially to the small numbers of source
samples used in the present paper.
The enhancement factor is actually equal to unity within $1\sigma$
level in the 1-Jy sample, despite the fact that the gravitational
lensing pair of Abell 2584 -- QSO 2319+272  (Stickel \& k\"uhr,
1992) is among the list of Table 1.
The situation is better for the 2-Jy sample,
which results in an enhancement of $q=1.51\sim3.02$ above $1\sigma$
level within $\theta=0.3^o$. This enhancement could be the result
of statistical gravitational lensing. If this is the case,
one may conclude that the associations of Abell clusters with
distant radio-bright quasars exist on scale of $\sim20$ arcminutes,
the average radius of Abell clusters (Brown \& Burns, 1991).  \\

Gravitational lensing fits (see Fig.2 and Fig.3) to both sets of data
have been done using singular isothermal models for matter
distributions in Abell clusters. The critical radius in the fit to
the result of the 2-Jy sample is $\theta_c=0.2^o$.  We can then find the
velocity dispersion ($\sigma_{||}$) to be required for the fitted
$\theta_c$ in terms of the definition of critical radius in a singular
isothermal model: $\theta_c=4\pi(\sigma_{||}/c)^2(D_{ds}/D_s)$,
where $D_{ds}$ and $D_s$ are the angular diameter distances of the
Abell clusters (lenses) and of the observer to the quasars (sources),
respectively. Taking the typical redshift of
Abell cluster to be 0.1 and that of distant radio-bright quasar to be 1,
and using the above fit of $\theta_c$ to the 2-Jy result (Fig.3),
we estimate  the velocity dispersion  $\sigma_{||}\approx5500$km/s.
Surprisingly, the required velocity dispersion
is  $\sim5$ times larger than  the actual values for Abell clusters,
making the gravitational lensing hypothesis be unlike for
the detected associations of Abell clusters  with
distant radio-bright  quasars on scale of $\sim20$ arcminutes.
In fact, the critical radius of a rich galaxy
cluster alone is smaller than 1 arcminute (Wu \& Hammer, 1994), in contrast to
the large values of $0.2^o$ in the above lensing model fit.  \\

Nevertheless, a similar feature appeared also in the study of quasar-galaxy
associations. Recall that to reach a modest enhancement factor of 2.5,
a velocity dispersion much larger than that measured for normal galaxies
($\sim500$ km/s) is needed in the lensing
model of galaxies (Webster, et al., 1988; Narayan, 1989). This apparent
dilemma had led one to conclude that either gravitational lensing hypothesis
could not produce the observed strong associations or the mass responsible
for the statistical lensing is not well modelled by a population of isolated
galactic masses.    The similar conclusions can now be drawn for the
possible associations of Abell clusters with distant radio-bright
quasars seen in the 2-Jy sample. We believe that if the 2-Jy radio
quasar-cluster associations above $1\sigma$ level are real,
the gravitational matter responsible for the
observed effect is not well represented by an isolated  cluster although
a single model with large velocity dispersion can
produce the $q\sim\theta$ curves shown in Fig.3. All the
matter inhomogeneities along the line of sight to the distant quasars
can contribute the gravitational effect and enhance the lensing
magnification. Therefore, we should consider the above  estimated
velocity dispersion, hence the mass,
to be the total mass integrated along the light path to the distant quasars.
In this case, the weak lenses of large-scale structures in the
universe should be taken into account (Bartelmann \& Schneider,
1993a,b,c).\\

Another possibility to get a high enhancement by an isolated cluster of
galaxies is that the slope of the number counts of the radio-bright
sources with flux is intrinsically steeper than the observed one.
Note that the present estimate of $q$ depends on the assumption that
the observed number counts of radio-bright sources are unaffected by
gravitational lensing. However, the $q\sim\theta$ fit to the 2-Jy
sample in Fig.3 can be equally made for a normal cluster by using a
steeper slope ($>1.62$) for background radio sources. This implies
that the ``lensing unaffected background hypothesis" is questionable.
The discovery by Rodrigues-Williams \& Hogan (1993)
reaches essentially the similar conclusion.\\

The associations of distant radio-bright quasars with foreground galaxy
clusters we have detected are not as significant as
the ones for distant optical-bright quasars ($B\leq18.5$) (Rodrigues-Williams
\& Hogan, 1993).  This arises probably due to different sizes of the two
kinds of radiations associated with quasars. Quasar can be regarded as
a pointlike source in optical wavelength while its radio emission stems
from an extended region. For a disk source with radius of $R_s$ the
maximum magnification by an isothermal sphere as lens is
$A_{max}=1+116.35(\sqrt{1+z}/(1+z))(\theta_c/10")
(R_s/10{\rm kpc})^{-1}\;h_{50}^{-1}$
(Wu \& Hammer, 1994), which prevents from reaching a relatively high
magnification. The optical continuum emitting region in quasar
estimated from its variability
is of $\sim10^{-3}$ pc and therefore, very large magnification appears
to be possible.  Hence, optically selected distant quasars may show much
stronger associations with foreground galaxy clusters than radio
selected ones. This explains the difference of the significances in the two
searches for distant bright quasar-cluster associations. \\

\begin{Large}
\begin{center}
{\bf 5. Conclusions}\\
\end{center}
\end{Large}
\bigskip

The strong associations between  distant ($z>0.5$)
radio-bright ($\geq1$Jy at 5Ghz and $\geq2$Jy at 2GHz)
quasars and Abell clusters have not been seen,
which  have probably arised from the scarcity of the source samples
used in the present paper. Nevertheless, above $1\sigma$ level the 2-Jy
radio sample does show an association with the foreground Abell clusters
on scale of $\sim20$ arcminutes. 9 Abell clusters have been detected
within $0.3^o$ of the 103 quasars in the 2-Jy sample while only 4
are expected to see if Abell clusters distribute randomly on the
sky. Gravitational lensing hypothesis can very well fit to the enhancement
variations with the search range in the 2-Jy sample  if (1) a considerably
large mass is assumed to associate with Abell clusters, in the sense that
the detected associations might be the result of weak lensing effect of
clusters of galaxies tracing large-scale structures in the universe,
and/or (2)the intrinsic radio-bright sources have a steeper slope
with flux than the actually observed one, in the sense that the
``lensing unaffected background hypothesis"  is questionable.\\

Following the discovery of quasar-galaxy associations, including optical
quasar-galaxy, radio quasar-galaxy and radio quasar-IRAS galaxy,   and
of optical bright quasar-cluster associations,
we have extended the searches for the associations to
the radio-bright quasars and the clusters of galaxies.
Unfortunately, the small samples of background sources have
led to some large statistical fluctuations in the present searches.
A preliminary result is that there may exist the associations
between the high redshift radio-bright quasars and the low redshift
clusters of galaxies on scale of a typical size of the Abell cluster.
If this is the case, our result confirms the recent report by
Rodrigues-Williams (1993) who have found a significant association
($4.7\sigma$ level) between the foreground Zwicky clusters and the
distant quasars. We are planning to extend the current work to involve the
distant 3CR galaxies and to statistically search for the possible
associations of the high-redshift 3CR galaxies with clusters claimed by
Hammer \& Le F\`evre (1990). The study of the associations of distant
objects with clusters of galaxies can provide very useful information
about the matter distribution on large-scales in the universe and has a
strong impact on gravitational lensing hypothesis.  \\

\bigskip

{\noindent}{\bf ACKNOWLEDGMENTS}\\
We are grateful to Fran\c{c}ois Hammer for reading the manuscript and
for valuable comments. WXP thanks CNRS and WONG for financial support.

\begin{large}
\begin{center}
{\bf References}\\
\end{center}
\end{large}

\bigskip
{\noindent} Abell G. O., Corwin J{\small R}. H., Olowin R. P., 1989,
           ApJS, 70, 1

{\noindent} Bartelmann M., Schneider P., 1991, A\&A, 248, 349

{\noindent} Bartelmann M., Schneider P., 1993a, A\&A, 268, 1

{\noindent} Bartelmann M., Schneider P., 1993b, A\&A, 271, 421

{\noindent} Bartelmann M., Schneider P., 1993c, A\&A, submitted

{\noindent} Bonnet H., Fort B., Kneib J.-P., Mellier Y., Soucail, G.,
           1993, A\&A, 280, L7

{\noindent} Brown D. L., Burns J. O., 1991, AJ, 102, 1917

{\noindent} Dunlop J. S., Peacock J. A., 1990, MNRAS, 247, 19

{\noindent} Gott III J. R., Gunn J. E., 1974, ApJ, 190, L105

{\noindent} Fugmann W., 1990, A\&A, 204, 73

{\noindent} Fugmann W., 1990, A\&A, 240, 11

{\noindent} Hammer F., Le F\`evre O., 1990, ApJ, 357, 38

{\noindent} Hewitt, A., Burbidge G., 1993, ApJS, 87, 451

{\noindent} K\"uhr H., Witzel A., Pauliny-Toth I. I. K., Nauber U.,
           1981, A\&AS, 45, 367

{\noindent} Lynds R., Petrosian V., 1986, BAAS, 18, 1014

{\noindent} Narayan R., 1989, ApJ, 339, L53

{\noindent} Robertson J. G., Roach G. J., 1990, MNRAS, 247, 387

{\noindent} Rodrigues-Williams L. L., Hogan C. J., 1993, BAAS, 25, 794

{\noindent} Soucail G., Mellier Y., Fort B., Picat J. P., 1987, A\&A, 172, L14

{\noindent} Stickel M., Fried J. W., k\"uhr H., 1989, A\&AS, 80., 103

{\noindent} Stickel M., K\"uhr H., 1992, A\&A, 264, 68

{\noindent} Stickel M., K\"uhr H., 1993a, A\&AS, 100, 395

{\noindent} Stickel M., K\"uhr H., 1993b, A\&AS, 101, 521

{\noindent} Stickel M., K\"uhr H., 1993c, A\&AS, in press

{\noindent} Stickel M., K\"uhr H., Fried J. W., 1993, A\&AS, 97, 483

{\noindent} Struble M. F., Rood H. J., 1991, ApJS, 77, 363

{\noindent} Tyson J. A., 1986, AJ, 92, 691

{\noindent} Unewisse A. M., Hunstead R. W., 1991, Proc. ASA, 9, 100

{\noindent} Wall J. V., Peacock J. A., 1985, MNRAS, 216, 173

{\noindent} Webster R. L., Hewitt P. C., Harding M. E., Wegner G. A.,
	   1988, Nat, 336, 358

{\noindent} Wu X. P., 1994, A\&A, 286, 748

{\noindent} Wu X. P., Hammer F., 1993, MNRAS, 262, 187

{\noindent} Wu X. P., Hammer F., 1994, A\&A, in press

\newpage

\begin{large}
\begin{center}
{\bf Figure Captions}\\
\end{center}
\end{large}

\bigskip
\bigskip
\bigskip
\bigskip
\bigskip

{\noindent}{\it Figure 1} ~~Enhancement $q$ of the radio-bright sources
versus gravitational magnification $A$.\\%

{\noindent}{\it Figure 2} ~~Variations of number excess of Abell clusters
around distant 1-Jy radio quasars.  The data have been normalized at
$3^o$. The fit to a singular isothermal sphere as lensing deflector
is shown using a critical radius of $0.1^o$.\\

{\noindent}{\it Figure 3} ~~The same as Fig.2 but for the 2 Jy sample
and a critical radius of $0.2^o$.\\

\newpage

\begin{center}

{\bf Table 1} Abell clusters detected within a search range\\
 of $0.3^o$ of the 1-Jy distant radio quasars\\

\bigskip
\bigskip

\begin{tiny}

\begin{tabular}{l|l|l|c||c|c|c|c|c|c}
\hline
{\small \null~~quasar} & {\small ~RA ~(1950)}  & {\small ~Dec~ (1950)} & z &
{\small Abell} &
 {\small RA (1950)} & {\small Dec (1950)} &  z & {\small richness} &
{\small separation} ($^o$)\\
\hline
0426$-$380  & 04 26 54.74 & $-$38  02 52.4 & 1.030 &
  3259 & 04 27.0 &$-$38 13 & (.1425)    & 3 & 0.17\\
0809+48 &  08  09 59.42 & +48 22  07.2 &0.871&
   637 & 08 11.4 &+48 32 & (.1485)    &   0& 0.29\\
0954+55  &  09 54 14.34&+55 37 16.6 &0.901&
   899&  09 54.9& +55 31&  (.1397) &   1& 0.14\\
1055+01&10 55 55.31&   +01 50  03.7 &0.888&
  1139& 10 55.5 &+01 46& .0397& 0 &0.13\\
1127$-$14  & 11 27 35.68& $-$14 32 54.8 &1.187&
  1285& 11 27.9 &$-$14 17 &.1010 &   1 & 0.28\\
1148$-$00 & 11 48 10.13& $-$00  07 13.2 &1.982&
  1392 &11 48.0 &$-$00 18    &.1382& 0& 0.18\\
1624+41 &       16 24 18.21 & +41 41 23.3 &2.550&
  2196 &16 25.7 &+41 36 &  .1332& 0&  0.28\\
2052$-$47 &   20 52 50.50& $-$47 26 19.6 &1.489&
  3720 &20 53.2& $-$47 13& (.1987)    &1& 0.23\\
2131$-$021 &      21 31 35.34& $-$02 06 40.8& 0.557&
  2353& 21 31.8 &$-$01 49& (.1078) &1&  0.30\\
2211$-$17 &22 11 42.51& $-$17 16 33.7& 2.153&
  3847& 22 11.8 &$-$17 16 & (.1500)     &0& 0.024\\
2319+272 &      23 19 31.99 & +27 16 19.1& 1.253&
  2584& 23 19.6& +27 17&  .1184& 0 &0.019\\
\hline
\end{tabular}
\end{tiny}
\end{center}

\newpage
\begin{center}
{\bf Table 2}  Number of the detected Abell clusters around\\
      distant radio-bright quasars in the 1 Jy and 2 Jy samples\\

\bigskip
\bigskip

\begin{tabular}{c|c|c|c|c}     \hline
range $\theta$ &
\multicolumn{2}{c|}{ N($\leq\theta)$} &
\multicolumn{2}{c}{$q$ (normalized at $3^o$)}\\
\cline{2-5}
(degree) & ~~~1 Jy~~~  & ~~~2 Jy~~~  & ~~~~1 Jy~~~~ & ~~~~2 Jy~~~~\\
\hline
0.2 &   6 &   4 &   1.63${\pm0.69}$ &      \\
0.3 &  11 &   9 &   1.33${\pm0.42}$  & 2.16${\pm1.75}$  \\
0.4 &  12 &  11 &   0.82${\pm0.24}$  & 1.48${\pm0.47}$  \\
0.5 &  18 &  14 &   0.78${\pm0.19}$  & 1.21${\pm0.34}$  \\
0.6 &  32 &  22 &   0.97${\pm0.18}$  & 1.32${\pm0.29}$  \\
0.7 &  45 &  28 &   1.00${\pm0.15}$  & 1.23${\pm0.24}$  \\
0.8 &  56 &  33 &   0.95${\pm0.13}$  & 1.11${\pm0.20}$  \\
0.9 &  77 &  41 &   1.03${\pm0.12}$  & 1.09${\pm0.18}$  \\
1.0 &  92 &  46 &   1.00${\pm0.11}$  & 0.99${\pm0.15}$  \\
1.5 & 198 & 110 &   0.96${\pm0.07}$  & 1.06${\pm0.11}$  \\
2.0 & 368 & 185 &   1.00${\pm0.05}$  & 1.00${\pm0.08}$ \\  \hline
\end{tabular}

\end{center}
\end{document}